\documentclass[aps,pra,twocolumn,amsmath,amssymb,amsfonts,floatfix,notitlepage,nofootinbib,superscriptaddress,showpacs]{revtex4-2}
\usepackage{times,graphics,graphicx,bm,bbm,color,eurosym,xurl,dsfont,mathtools,physics,placeins,diagbox,multirow,booktabs}
\usepackage{hyperref}
\hypersetup{breaklinks=true}
\hypersetup{colorlinks,linkcolor={blue},citecolor={blue},urlcolor= {blue}}
\usepackage[T1]{fontenc} 
\usepackage{newtxmath} 
\usepackage{newtxtext} 
\DeclareFontFamily{OT1}{pzc}{}
\DeclareFontShape{OT1}{pzc}{m}{it}%
              {<-> s * [1.25] pzcmi7t}{}
\DeclareMathAlphabet{\mathpzc}{OT1}{pzc}%
                                 {m}{it}

\DeclareUnicodeCharacter{200E}{}

\begin{document}

\title{Feature- and process-based optimal control of quantum dynamics}

\author{M. Farnia}
\affiliation{Department of Physics, Sharif University of Technology, Tehran 14588, Iran}

\author{V. Rezvani}
\email{vahid_rezvani@iau.ac.ir}
\affiliation{Department of Fundamental Science, Has.C., Islamic Azad University, Hashtgerd, Iran}

\author{A. T. Rezakhani}
\affiliation{Department of Physics, Sharif University of Technology, Tehran 14588, Iran}

\begin{abstract}
Preparing desired quantum states and quantum operations (processes) is essential for numerous tasks in quantum computation. Several approaches have been developed for optimal control of quantum states, whereas optimal strategies for preparation of a given quantum process have remained fairly less explored. For some applications, rather than a specific desired state or process, it may suffice to obtain states or processes with specific desired \textit{features} such as (high) coherence or purity. In such cases, fidelity-based measures alone are inadequate for evaluating the performance of a quantum control strategy, and hence proper feature-based figures of merit should be employed. Here we develop a feature-based optimal coherent control formalism for open quantum processes under Markovian evolutions which demonstrate high coherence and purity features. In particular, we observe that performance of the feature-based Krotov optimization algorithm can be improved by choosing educated initial guess fields. In addition, in a model for a qutrit Rydberg system, it is shown that the convex overlap-based fidelity outperforms other process-based as well as state-based fidelities. This analysis underscores the utility of feature-based optimal control strategies for quantum processes.
\end{abstract}
\date{\today}
\maketitle

\section{Introduction}

In the quantum world, some operations can happen faster, more accurate, and more intricate than the classical world. This relative advantage is due to unique quantum resources such as entanglement and quantum coherence. However, one of the major challenges in quantum technologies is to protect or improve these resources under system-environment effects \cite{breuer_theory_2002, rivas_2012}. To this end, several strategies have been developed for active and passive control  \cite{nielsen_quantum_2010, book:QEC, viola_dynamical_1999, Khodjasteh, Viola-review, lloyd_coherent_2000, lidar_review_2012}. 

In optimal (coherent) control theory \cite{book:D'Alessandro, book:q.control, Jirari_Optimal_Control, Rezvani-thesis}, as an active strategy where control fields are applied on an open quantum system, the goal is to improve performance of the system for a specific objective, such as preparation of a given target state or gate at a specific final time or for the whole duration of the dynamics. Performance is often characterized with some appropriate \textit{objective} functional, which compares how close to the target state or gate the strategy has taken the system. The objective functional needs to be optimized under specific physical constraints such as a given evolution scenario under environmental effects. 
When we aim to characterize and control the quantum process acting on an open system, this information is unambiguously included in its ``process matrix'' \cite{nielsen_quantum_2010, PhysRevA.71.062310}. This matrix can be conveniently measured by quantum process tomography methods in the lab \cite{Chuang01111997, PhysRevLett.78.390, PhysRevLett.86.4195, 7f797eca63814c35b36787ec167bc2b4, Mohseni_2006, PhysRevA.77.032322, PhysRevLett.100.190403}, and its dynamics can be described by proper master equations \cite{mohseni2009equation, rezvani2021coherent, AGK-ATR}. In addition, some quantum \textit{features} of a dynamics, such as coherence \cite{datta2018coherence, xu2019coherence} and purity \cite{kong2016state}, can also be described by its process matrix. Having such utilities can make process matrices particularly useful for designing optimal control schemes to directly manipulate the open-system process \cite{rezvani2021coherent}.

For some practical applications, rather than generating a particular target process, it may suffice to reach a process which has a particular feature, e.g., a high coherence or purity. To analyze such \textit{feature-based} control problems, common quantum fidelity functionals may appear inadequate. In fact, it has been known that two states or processes with high quantum fidelities do not necessarily have close quantum features \cite{PhysRevA.87.052136, bina2014drawbacks}. As such, one shall also need to use appropriate feature-based objective functionals for this type of optimal control problems. Since target processes with some optimal value of a particular feature need not be a unique process, optimization can be more sophisticated than typical problems. 

Here we study a \textit{feature-based} optimal control scheme whose objective is to optimally generate processes with particular features, rather than generate a particular target process. We consider an open system under a Markovian evolution within an environment, where we only have control over the system Hamiltonian (for a precursor study in the context of closed quantum systems see Ref. \cite{PhysRevA.70.052313}). 
We employ the Krotov optimization method \cite{Konnov-krotov, krotov1996, reich2012monotonically}, which is a powerful, monotonically convergent iterative algorithm. Performance of this algorithm may depend on the topology of objective functionals \cite{Werschnik_2007, Floether_2012, PhysRevA.91.043401} and hence the choice of initial guess fields. To improve this performance by \textit{educated} initial fields, we first optimize the same functional or a different one (such as the process fidelity) at an intermediate time and then use this pre-optimized field as the initial guess field for the main optimization process.

The structure of this paper is as follows. In Sec. \ref{sec.2}, we state the dynamical optimization problem for processes and review some preliminary materials. Several fidelities and features, including coherence and purity of a dynamics, are considered in Sec. \ref{sec.4}. Section \ref{sec.5} illustrates the feature-based optimization for a Rydberg qutrit system. Section \ref{sec.6} concludes the paper.  

\section{Optimal control of quantum dynamics}
\label{sec.2}

We consider dynamics of a specific open quantum system ($S$), which interacts with its surrounding environment ($E$). The main goal of a dynamical optimization problem is to find the optimal values of a total objective functional as  
\begin{equation}
\mathpzc{J} = \mathpzc{F}\big(\chi(t_{\mathrm{f}}),t_{\mathrm{f}}\big) + \mathpzc{J}_{d}[\chi(t),\boldsymbol{\epsilon}(t)].
\label{eq.1}
\end{equation}
Here $\mathpzc{F}$ is the \textit{final-time} functional, with $\chi(t_{\mathrm{f}})$ as an appropriate dynamical variable of the problem which represents the dynamics at the final time $t_{\mathrm{f}}$. For example, $\mathpzc{F}$ can be chosen to be a fidelity measure between the actual dynamics $\chi(t_{\mathrm{f}})$ at  $t_{\mathrm{f}}$ with the desired target dynamics $\Xi$. Later we discuss several fidelity measures between two dynamics, which are useful for our purposes in this paper. The dynamics-dependent functional $\mathpzc{J}_{d}$ depends on the dynamical variable $\chi(t)$ and external fields $\boldsymbol{\epsilon}(t) = \{\epsilon_{m}(t) \}$ at $t\in [0, t_{\mathrm{f}})$. In practical applications, the control fields $\boldsymbol{\epsilon}(t)$ act as adjustable knobs to steer the dynamics of the system. The functional $\mathpzc{J}_{d}$ can include all experimental constraints on the external fields, such as energy limitations imposed by the lab setup. 

\subsection{Dynamical variable for open quantum system}
\label{sec.2-subsec.1}

To define a dynamical variable $\chi(t)$, we assume the initial state of the total system has a tensor-product form, $\varrho(t_{0})=\varrho_{S}(t_{0})\otimes\varrho_{E}(t_{0})$, where $\varrho_{S}(t_{0})$ ($\varrho_{E}(t_{0})$) is the initial state of the quantum system (environment). Hence the state of the system $S$ at later times $t\geqslant t_{0}$ is given by a completely positive and trace-preserving linear map $\mathpzc{E}_{(t,t_{0})}$ through $\varrho_{S}(t)=\mathpzc{E}_{(t,t_{0})}[\varrho_{S}(t_{0})]$ where 
\begin{equation}
\mathpzc{E}_{(t,t_{0})}[\cdot]=\textstyle{\sum}_{\lambda,\mu =1}^{N^{2}}\chi_{\lambda\mu}(t,t_{0})\, C_{\lambda}\cdot C_{\mu}^{\dag},
\label{eq.2}
\end{equation}
and $N$ is the dimension of the Hilbert space of the system \cite{breuer_theory_2002, rivas_2012, lidar2019lecture}. The operator set $\{C_{\lambda}\}_{\lambda=1}^{N^{2}}$ constitutes a fixed orthonormal operator basis for the $N^{2}$-dimensional Liouville space of $S$ with the orthonormality condition $\mathrm{Tr}[C_{\lambda}^{\dag}C_{\mu}]=\delta_{\lambda\mu}$. For example, this operator basis can be the logical operator basis $\{\tilde{C}_{(i,j)}=\vert i\rangle\langle j\vert\}_{i,j=1}^{N}$ with $\vert i\rangle$ as the computational basis or generalized Gell-Mann basis matrices \cite{Bertlmann_2008}. Equation \eqref{eq.2} is the Kraus representation of the dynamical map $\mathpzc{E}_{(t,t_{0})}$ in the $C_{\lambda}$ basis \cite{nielsen_quantum_2010}. The coefficients of the expansion \eqref{eq.2} form a matrix $\chi$, referred to as the \textit{process matrix}, which is defined as
\begin{equation}
\begin{split}
&\chi(t,t_{0})=\mathpzc{B}^{\dag}(t,t_{0})\mathpzc{B}(t,t_{0}),
\\
&\mathpzc{B}_{(i,j),\lambda}(t,t_{0})=\sqrt{r_{i}}\,\mathrm{Tr}[\langle b_{i}\vert U^{\dag}(t,t_{0})\vert b_{j}\rangle C_{\lambda}],
\end{split}
\label{eq.4}
\end{equation}
where the set $\{ r_{i};\vert b_{i}\rangle \}$ consists of all eigenvalues and corresponding eigenvectors of the initial state of the environment $\varrho_{E}(t_{0})$. In Eq. \eqref{eq.4}, the total unitary operator $U(t,t_{0})$ is generated by the total Hamiltonian $H(t)=H_{S}+V_{\mathrm{field}}(t)+H_{E}+H_{SE}$, where $H_{S}$ ($H_{E}$) is the free Hamiltonian of the open system (environment), $H_{SE}$ is the system-environment interaction, and the Hamiltonian $V_{\mathrm{field}}(t)=\textstyle{\sum}_{i}\epsilon_{m}(t)H_{m}$ ($H_{SE}$) describes the action of external control fields on the system, with the control Hamiltonians $\{ H_{m}\}$ determined by accessible control scenarios in the lab. Equation \eqref{eq.4} ensures that the $\chi$ matrix embodies all dynamical information of the open system $S$. This matrix then is a suitable candidate for the dynamical variable in optimal control and has a pivotal role in our scheme. Hereafter, for brevity we represent the $\chi(t,t_{0})$ simply with $\chi(t)$. 

The trace-preserving condition of the dynamical map $\mathpzc{E}_{(t,t_{0})}$ implies that $\mathrm{Tr}[\chi(t)]=N$ for all times $t\geqslant t_{0}$. From Eq. \eqref{eq.4}, it is evident that $\chi(t)$ is positive semidefinite at all times. In addition, there are plenty of quantum process tomography techniques which aim to determine process matrices through direct \cite{Mohseni_2006, Mohseni_2008} or indirect \cite{PhysRevLett.78.390, 7f797eca63814c35b36787ec167bc2b4} methods in the lab.

\begin{figure}[tp]
\includegraphics[width=.8\linewidth]{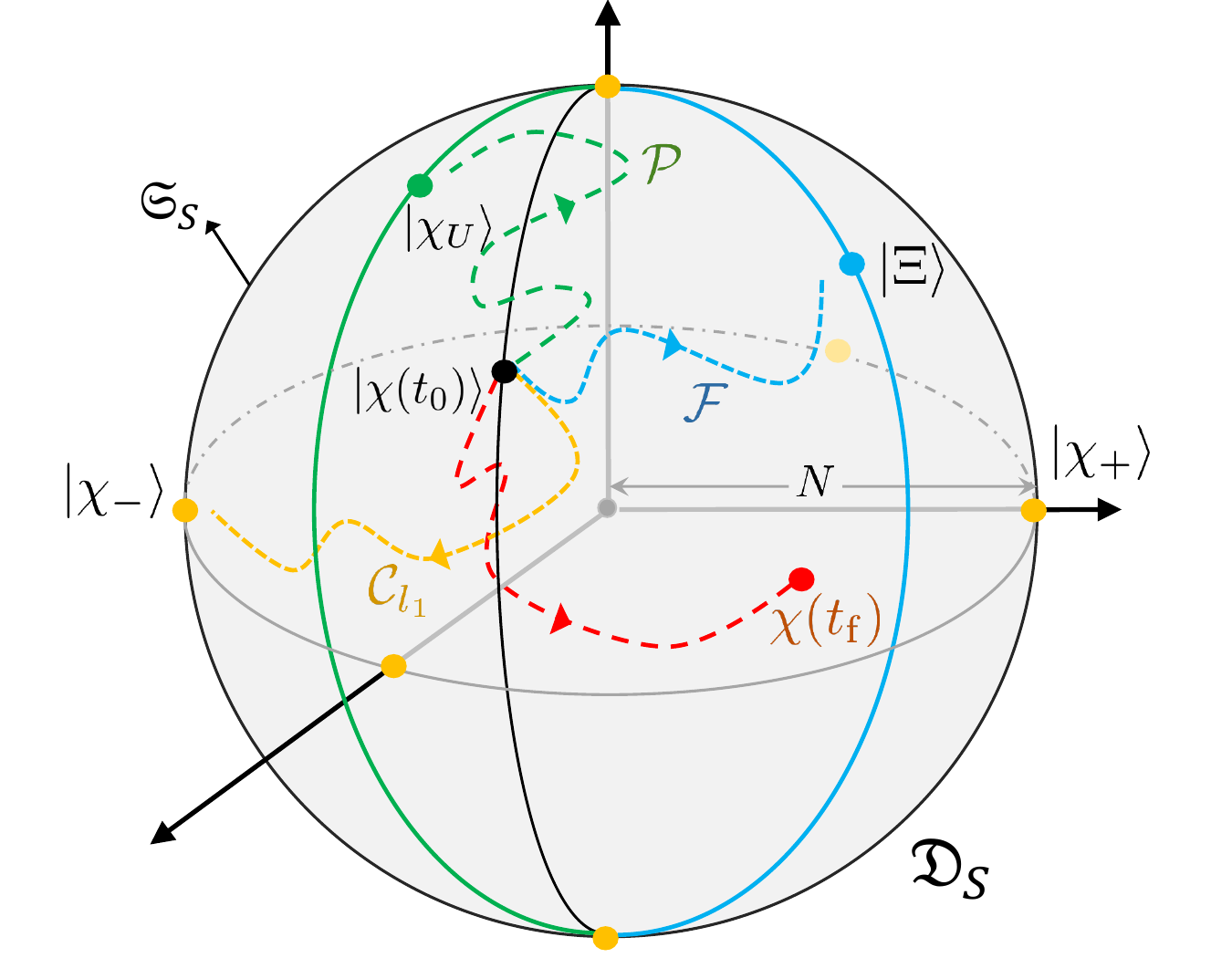}
\caption{Schematic of the process-based optimal control scenario. The $(N^{4}-1)$-dimensional hypersphere of radius $N$ encompasses all process matrices constituting the dynamical space $\mathfrak{D}_{S}$ of the system. All unitary dynamics lie on the surface (shown by $\mathfrak{S}_{S}$) of this hypersphere. The red dashed curve shows time evolution of the dynamics generated by $H=H_{S}+H_{E}+H_{SE}$. The initial unitary and final dynamics have been indicated by $\ket{\chi(t_{0})}$ and $\chi(t_{\mathrm{f}})$, respectively. The blue dashed curve shows time evolution of the process matrix obtained from optimizing the fidelity $\mathpzc{F}$ between the final dynamics and the desired unitary gate $\ket{\Xi}$. The green (orange) dashed curve represents the evolution trajectory of the dynamics resulting from optimizing purity (coherence) of the dynamics at $t_{\mathrm{f}}$. The dynamics with maximal coherence are indicated by orange dots. The $\ket{\chi_{U}}$ is an arbitrary unitary dynamics. Here $\ket{\cdot}$ shows a pure state corresponding to an arbitrary unitary dynamics, while $|\cdot \rangle \hskip-0.7mm\rangle$ represents the vectorized form of a process matrix. See the main text.}
\label{fig:ContSce}
\end{figure}

The $\chi$ matrix has also a clear connection with the Choi-Jamiolkowski isomorphism \cite{JAMIOLKOWSKI1972275, CHOI1975285}, a one-to-one correspondence between the dynamical map $\mathpzc{E}_{(t,t_{0})}$ and a density matrix in the $N^{2}$-dimensional Hilbert space of the open system and a similar ancillary system given by $\varrho_{\mathpzc{E}}(t) = ( \mathpzc{E}_{(t,t_{0})} \otimes \mathbbmss{I}_{S})[\vert\phi^{+}\rangle\langle\phi^{+}\vert]$, where $\vert\phi_{+}\rangle=\sum_{i}|i\rangle|i\rangle/\sqrt{N}$. This implies that $\tilde{\chi}(t)$ in the logical basis $\{\tilde{C}_{(i,j)}\}$ is related to the density matrix $\varrho_{\mathpzc{E}}(t)$ as $\varrho_{\mathpzc{E}}(t) = (1/N)\tilde{\chi}(t)$. The process matrices $\tilde{\chi}(t)$ and $\chi(t)$ in another basis $\{ C_{\alpha}\}$ are connected by
\begin{equation}
\Tilde{\chi}(t)=\mathpzc{U}^{\dag}\chi(t)\mathpzc{U},
\label{eq.5}
\end{equation} 
where $\mathpzc{U}_{\alpha,(i,j)}=\mathrm{Tr}[C_{\alpha}^{\dag}\Tilde{C}_{(i,j)}]$ is a unitary operator. In addition, we have
\begin{equation}
\mathrm{Tr}[\chi^{2}(t)]\leqslant N^{2},
\label{eq.6}
\end{equation}
where the equality holds only for unitary dynamical operators $U$, for which $\chi$ matrix $[\chi_{U}]_{\lambda,\mu} = \mathrm{Tr}[UC_{\lambda}^{\dag}]\, \mathrm{Tr}[UC_{\mu}^{\dag}]^{*}$. Thus, one can consider unitary dynamics as a \textit{pure} state (shown by $\vert\chi_{U}\rangle$) belonging to an $N^{4}$-dimensional space $\mathfrak{D}_{S}$ including all process matrices. Concluding these considerations, we can represent the dynamical space $\mathfrak{D}_{S}$ geometrically as an $(N^{4}-1)$-dimensional hypersphere of radius $N$ where all unitary dynamics lie on the surface $\mathfrak{S}_{U}$ (Fig. \ref{fig:ContSce}). This enables us to assign a Bloch representation as
\begin{equation}
\chi(t)=(1/N)\mathbbmss{I}_{N^{2}}+ \textstyle{\sum}_{\alpha=1}^{N^{4}-1}r_{\alpha}(t) M_{\alpha},
\label{eq.6-1}
\end{equation}
where $\boldsymbol{r}(t)=\{r_{\alpha}(t)\}$ is the generalized Bloch vector of dimension $N^{4}-1$ and $\{ M_{\alpha} \}_{\alpha = 1}^{N^{4}}$ is an orthonormal operator basis for an $N^{4}$-dimensional vector space such that $\mathrm{Tr}[M_{\alpha}] = 0$ for $\alpha\in\{1,2,\ldots,N^{4}-1 \}$ and $M_{N^{4}} = (1/N)\mathbbmss{I}_{N^{2}}$. 

\subsection{Markovian master equation for process matrices}
\label{sec.2-subsec.2}

In optimal control theory, it is necessary to consider the time evolution of the dynamical variable. 
Since the process matrix $\chi(t)$ in our dynamical variable, hence we need to study its time evolution. Under \textit{Markovian} assumptions \cite{breuer_theory_2002, lidar2019lecture}, the state of the system evolves as
\begin{align}
\frac{d\varrho_{S}(t)}{dt} =& -\frac{i}{\hbar} [H_{S}(t),\varrho_{S}(t)] + \textstyle{\sum_{\alpha=1}^{N^2-1}}\gamma_{\alpha}\big(L_{\alpha}\varrho_{S}(t)L^{\dag}_{\alpha}\nonumber\\
&- (1/2)\{L^{\dag}_{\alpha}L_{\alpha},\varrho_{S}(t)\} \big),
\end{align}
where $\gamma_{\alpha}\geqslant 0$ are the quantum jump rates and $L_{\alpha}$s are the jump operators. It has been shown in Ref. \cite{rezvani2021coherent} that the previous Lindblad equation translates as the following for the process matrix:
\begin{align}
\frac{d\chi(t)}{dt}=&-\frac{i}{\hbar}[\mathsf{H}_{S}(t),\chi(t)] + \textstyle{\sum_{\alpha=1}^{N^{2}-1}} \gamma_{\alpha}
\big{(}\mathsf{L}_{\alpha}\chi(t)\mathsf{L}_{\alpha}^{\dag}\nonumber\\ 
&- (1/2)\{\mathsf{L}_{\alpha}^{\dag}\mathsf{L}_{\alpha}, \chi(t)\}\big{)}=: -\frac{i}{\hbar} \mathpzc{K}_{\;\;\boldsymbol{\epsilon}}\big{(}\chi(t)\big{)},
\label{eq.8}
\end{align}
where $\chi_{\lambda\mu}(t_{0})=N\delta_{\lambda N^{2}}\delta_{\mu N^{2}}$ and $[\mathsf{Y}]_{\lambda,\mu} = \mathrm{Tr}[C_{\lambda}^{\dag}Y C_{\mu}]$ with $Y\in\{ H_{S}(t), L_{\alpha}\}$. We shall use a (basis-dependent) vectorized form of this dynamical equation in the following, which is defined by $X=\sum_{ij}X_{ij}|i\rangle \langle j|\to|X\rangle\hskip-0.7mm\rangle=\sum_{ij}X_{ij}|i\rangle|j \rangle$ and has the properties $XYZ\to(X\otimes Z^{T})|Y\rangle\hskip-0.7mm\rangle$ (with $T$ denoting transposition) and $\langle\hskip-0.7mm\langle X|Y\rangle\hskip-0.7mm\rangle=\mathrm{Tr}[X^{\dag}Y]$. In the vectorized dynamical equation we use the notation $\chi(t) \to |\chi(t)\rangle\hskip-0.7mm\rangle$ and $\mathpzc{K}_{\,\,\epsilon}\to\mathbbmss{K}_{\epsilon}$.

Now all necessary tools are in order for our open-system dynamical control: obtaining $\{\epsilon_{m} (t)\}$ for which the total functional \eqref{eq.1} is optimized provided that the dynamical variable $|\chi(t)\rangle\hskip-0.7mm\rangle$ evolves according to the vectorized form of Eq. \eqref{eq.8}. 

\subsection{Optimization algorithm}
\label{sec.3}

To solve our optimization problem, we can use any gradient-free or gradient-based algorithm. Example of gradient-based algorithms are the Zhu-Rabitz algorithm \cite{zhu1998rapid, zhu1998rapidly}, gradient ascent pulse engineering \cite{KHANEJA2005296}, Newton and quasi-Newton methods \cite{gill1972quasi, PhysRevA.83.053426}, and the Krotov algorithm \cite{ Konnov-krotov, krotov1996, reich2012monotonically}. In tis paper we use the Krotov method. Although some optimization algorithms such as the Zhu-Rabitz method are suitable for specific problems, e.g., those with convex final-time functionals, the Krotov  algorithm applies to a wider class of problems, including those with nonconvex functionals and nonlinear master equations \cite{sklarz2002loading, reich2012monotonically}. Having a total objective functional and a dynamical equation, this method updates the initial guess fields iteratively and in a monotonically convergent fashion. In the following, we reproduce and modify necessary steps on this algorithm based on Ref. \cite{rezvani2021coherent}.

The input of the $n$th iteration of the Krotov algorithm is the external fields $\boldsymbol{\epsilon}^{(n-1)}(t)$ and the process matrix $|\chi^{(n-1)}(t)\rangle \hskip-0.7mm\rangle$ obtained from the $(n-1)$th iteration ($n\in\{ 1,2,3,\ldots\}$), with $\bm{\epsilon}^{(0)}(t)$ being the initial guess fields. This dynamics depends on $\bm{\epsilon}^{(n-1)}(t)$ through the dynamical equation $d\vert\chi^{(n-1)}(t)\rangle\hskip-0.7mm\rangle/dt=-(i/\hbar)\mathbbmss{K}_{\bm{\epsilon}^{(n-1)}}\vert\chi^{(n-1)}(t)\rangle\hskip-0.7mm\rangle$ with the initial condition $\vert\chi^{(n-1)}(t_{0})\rangle\hskip-0.7mm\rangle_{\alpha}=N\delta_{\alpha N^{4}}$. At the first step of the $n$th iteration, we need to solve the following dynamical equation for an $N^{4}$-dimensional vector $\vert\Lambda (t)\rangle\hskip-0.7mm\rangle$:
\begin{align}
&\frac{d\vert\Lambda(t)\rangle\hskip-0.7mm\rangle}{dt}=-\frac{i}{\hbar}\mathbbmss{K}^{\dag}_{\bm{\epsilon}^{(n-1)}}\vert\Lambda(t)\rangle\hskip-0.7mm\rangle,\label{eq.10} \\
&\vert\Lambda(t_{\mathrm{f}})\rangle\hskip-0.7mm\rangle =-\Big( \frac{\partial\mathpzc{F}}{\partial \langle\hskip-0.7mm\langle\chi(t)\vert}\Big)\Big| _{\chi^{(n-1)}(t_{\mathrm{f}})}.
\label{eq.11}
\end{align}

For the second step of the $n$th iteration, it is necessary to choose a specific form for $\mathpzc{J}_{d}$ [Eq. \eqref{eq.1}]. This functional is often assumed to depend only on the control fields $\{\epsilon_{m} (t)\}$ as 
\begin{align}
\mathpzc{J}_{d}[\bm{\epsilon}(t)] = {\textstyle{\sum_{m}}} \int_{t_{0}}^{t_{\mathrm{f}}}dt\,\frac{w_{m}}{f_{m}(t)}\big(\epsilon_{m}(t)-\epsilon_{m}^{(\mathrm{ref})}(t) \big)^{2},
\label{eq.12}
\end{align}
where $0\leqslant w_{m}\leqslant 1$, and $f_{m}(t)$ and $\epsilon_{m}^{(\mathrm{ref})}(t)$ are the shape function and reference field for the control field $\epsilon_{m}(t)$, respectively. 
The update rule for the fields is obtained as \cite{rezvani2021coherent}
\begin{align}
\epsilon_{m}^{(n)}(t)=\, &\epsilon_{m}^{(\mathrm{ref})}(t)+\dfrac{f_{m}(t)}{ \hbar w_{m}}\Big\{\mathrm{Im}\langle\hskip-0.7mm\langle\Lambda(t)‎\vert\Big(\dfrac{\partial\mathbbmss{K}_{\bm{\epsilon}}}{\partial\epsilon_{m}}\Big)\Big| _{\bm{\epsilon}^{(n)}}\vert\chi^{(n)}(t)\rangle\hskip-0.7mm\rangle‎
‎\nonumber\\
‎\, &+\dfrac{\sigma}{2}\mathrm{Im}\langle\hskip-0.7mm\langle\Delta\chi(t)‎\vert\Big(\dfrac{\partial\mathbbmss{K}_{\bm{\epsilon}}}{\partial\epsilon_{m}}\Big)\Big| _{\bm{\epsilon}^{(n)}}\vert\chi^{(n)}(t)\rangle\hskip-0.7mm\rangle\Big\},
\label{eq.13}
\end{align}
where $\vert \Delta\chi^{(n)}(t)\rangle\hskip-0.7mm\rangle=\vert \chi^{(n)}(t)\rangle\hskip-0.7mm\rangle‎ -‎\vert \chi^{(n-1)}(t)\rangle\hskip-0.7mm\rangle$. This relation should be considered with the equation $d\vert\chi^{(n)}(t)\rangle\hskip-0.7mm\rangle/dt=-(i/\hbar)\mathbbmss{K}_{\bm{\epsilon}^{(n)}}\vert\chi^{(n)}(t)\rangle\hskip-0.7mm\rangle$ (with the boundary condition $\vert\chi^{(n)}(t_{0})\rangle\hskip-0.7mm\rangle_{\alpha}=N\delta_{\alpha N^{4}}$). The last term is of second order with respect to the $|\chi\rangle\hskip-0.7mm\rangle$, with
\begin{align}
&\sigma=-\Bar{A},\qquad‎\bar{A}=\mathrm{max}\{\zeta_{A},2A+\zeta_{A}\},\qquad\zeta_{A}\in\mathrm{I\! R}^{+}‎,\nonumber\\
&‎A=\dfrac{\Delta\mathpzc{F}^{(n)}+2\mathrm{Re}\langle\hskip-0.7mm\langle\Delta\chi^{(n)}(t_{\mathrm{f}})\vert\Lambda(t_{\mathrm{\mathrm{f}}})\rangle\hskip-0.7mm\rangle}{\langle\hskip-0.7mm\langle\Delta\chi^{(n)}(t_{\mathrm{f}})\vert\Delta\chi^{(n)}(t_{\mathrm{f}})\rangle\hskip-0.7mm\rangle}‎,
\label{eq.14}
\end{align}
‎where $\Delta\mathpzc{F}^{(n)}=\mathpzc{F}\big(\chi^{(n)}(t_{\mathrm{f}}),t_{\mathrm{f}}\big)-\mathpzc{F}\big( \chi^{(n-1)}(t_{\mathrm{f}}),t_{\mathrm{f}}\big)$‎.
Throughout this paper, we set $\epsilon_{m}^{(\mathrm{ref})}(t)=\epsilon_{m}^{(n-1)}(t)$. With this choice, near the convergence point the dynamics-dependent functional $\mathpzc{J}_{d}$ approaches zero and $\mathpzc{J}=\mathpzc{F}$. In addition, for the convex final-time-dependent functional, one can show $\Bar{A}=0$ by setting $\zeta_{A}=0$, that is, for this case the Krotov algorithm becomes first order with respect to $|\chi\rangle\hskip-0.7mm\rangle$.


\section{Process-based functionals}
\label{sec.4}

In this section, we recall two types of process-based (state-independent) final-time functionals $\mathpzc{F}$: fidelity measures and quantum features of a dynamics. Note that during optimization, we need to change $\mathpzc{F}\rightarrow -\mathpzc{F}$ to have a minimization problem so that we can use the Krotov algorithm.  

\subsection{Fidelity measures}
\label{sec.3-subsec.1}

One relevant option for $\mathpzc{F}$ is the process-based fidelity between the evolved process $\vert\chi(t_{\mathrm{f}})\rangle\hskip-0.7mm\rangle$ at a given final time $t_{\mathrm{f}}$ and a target dynamics $\vert\Xi\rangle\hskip-0.7mm\rangle$. In fact, here, we introduce some measures to quantify the degree of closeness between these dynamics in the space $\mathfrak{D}_{S}$. 
By extending arguments of Ref. \cite{basilewitsch2019quantum}, it is discerned that the minimum requirements for the fidelity functional $\mathpzc{F}(\chi_{1},\chi_{2})$ in the context of process-based optimal control (i.e., for all $\chi_{1},\chi_{2}\in\mathfrak{D}_{S}$) are as follows: (i) $\mathpzc{F}(\chi_{1}, \chi_{2}) \in \mathrm{I\!R}$, (ii) $\mathpzc{F}(\chi_{1}, \chi_{2}) = 1 \Leftrightarrow \chi_{1} = \chi_{2}$. 
In addition, the differentiability of the fidelity functional with respect to the control variables $\chi$ is essential for gradient-based algorithms such as the Krotov algorithm. 

Here we introduce several fidelities for process-based optimization applications. The first functional is the convex overlap-based fidelity defined as  
\begin{equation}
\mathpzc{F}_{\mathrm{c}}(\chi_{1}, \chi_{2}) = \langle\hskip-0.7mm\langle\chi_{1}\vert \chi_{2}\rangle\hskip-0.7mm\rangle/N^{2},
\label{eq.16}
\end{equation}
where $0\leqslant \mathpzc{F}_{\mathrm{c}}\leqslant 1$. It is straightforward to verify that $\mathpzc{F}_{\mathrm{c}}\big(\chi, \chi_{O}\big)= \mathpzc{F}_{\mathrm{Uhl}}(\chi, \chi_{O}) = \big( \mathrm{Tr}[ \sqrt{\sqrt{\chi_{O}} \chi \sqrt{\chi_{O}}}] \big)^{2}/N^{2}$ for the unitary gate $O$, where $\mathpzc{F}_{\mathrm{Uhl}}$ is the Uhlmann fidelity \cite{UHLMANN1976273,Jozsa01121994}. 
In addition, convexity of $\mathpzc{F}_{\mathrm{c}}$ guarantees that the optimal parameter ${A}=0$ [Eq. \eqref{eq.14}]. 

Another functional with the above conditions is the nonconvex overlap-based functional given by
\begin{align}
\mathpzc{F}_{\mathrm{nc}}(\chi_{1}, \chi_{2}) = \frac{\langle\hskip-0.7mm\langle\chi_{1}\vert \chi_{2}\rangle\hskip-0.7mm\rangle}{\sqrt{\langle\hskip-0.7mm\langle\chi_{1}\vert \chi_{1}\rangle\hskip-0.7mm\rangle \, \langle\hskip-0.7mm\langle\chi_{2}\vert \chi_{2}\rangle\hskip-0.7mm\rangle}}.
\label{eq.17}
\end{align}
The Cauchy-Schwarz inequality yields $N^{-2} \leqslant \mathpzc{F}_{\mathrm{nc}}\leqslant 1$. Nonconvexity of $-\mathpzc{F}_{\mathrm{nc}}$ implies that $A\neq 0$. 

The next functional we consider is the often-used (normalized) Hilbert-Schmidt fidelity, defined as
\begin{gather}
{\mathpzc{F}}_{\mathrm{HS}}(\chi_{1}, \chi_{2}) = 1- {\mathpzc{D}}_{\mathrm{HS}}(\chi_{1}, \chi_{2}), 
\label{eq.18}\\
\mathpzc{D}_{\mathrm{HS}}(\chi_{1}, \chi_{2}) = \langle\hskip-0.7mm\langle\chi_{1} - \chi_{2}\vert \chi_{1} - \chi_{2}\rangle\hskip-0.7mm\rangle/(2N^{2}). \nonumber
\end{gather}
From Eq. \eqref{eq.14}, one can see that $A=1/(2N^{2})$ for the $-{\mathpzc{F}}_{\mathrm{HS}}$ functional. According to Eq. \eqref{eq.6-1}, we can rewrite this fidelity functional as ${\mathpzc{F}}_{\mathrm{HS}}(\chi_{1}, \chi_{2})=1- \Vert\boldsymbol{r}_{1}-\boldsymbol{r}_{2}\Vert^{2}/(2N^{2})$, where $\boldsymbol{r}_{1}$ ($\boldsymbol{r}_{2}$) are the Bloch representation of $\chi_{1}$ ($\chi_{2}$). 

The length and direction contributions of the Bloch vectors $\boldsymbol{r}_{1}$ and $\boldsymbol{r}_{2}$ are then intertwined. Inspired by Ref. \cite{basilewitsch2019quantum}, to split these contributions we define the geometric functional
\begin{gather}
\mathpzc{F}_{\mathrm{geo}}(\chi_{1}, \chi_{2}) = 1- \mathpzc{D}_{\mathrm{geo}}(\chi_{1}, \chi_{2}),
\label{eq.19}\\
\mathpzc{D}_{\mathrm{geo}}(\chi_{1}, \chi_{2}) = w_{1} \mathpzc{D}_{\mathrm{angle}}(\chi_{1}, \chi_{2}) + w_{2}
\mathpzc{D}_{\mathrm{length}}(\chi_{1}, \chi_{2}),
\nonumber
\end{gather}
where $0\leqslant w_{i}\leqslant 1$ and $\textstyle{\sum}_{i=1}^{2}w_{i}=1$. Here the length and direction contributions of the Bloch vectors are defined as
\begin{align*}
&\mathpzc{D}_{\mathrm{length}} = \frac{1}{N^{2}-1} \big(\sqrt{d_{11}} - \sqrt{d_{22}}\big)^{2},\\
&\mathpzc{D}_{\mathrm{angle}} = \frac{1}{\pi^{2}} \arccos^{2} \left(\frac{d_{12}}{\sqrt{d_{11} d_{22}}} \right),
\end{align*}
where $d_{ij} = \langle\hskip-0.7mm\langle\chi_{i}\vert \chi_{j}\rangle\hskip-0.7mm\rangle-1 $. It is evident that $-\mathpzc{F}_{\mathrm{geo}}$ is nonconvex and hence $A\neq 0$. This splitting provides some flexibility in determining the role of different contributions during an optimization algorithm.

The last fidelity measure we consider is the following \textit{state-dependent} fidelity $\mathpzc{F}_{\mathrm{S}}$ to simulate a target unitary gate $O$:
\begin{equation}
\mathpzc{F}_{\mathrm{S}}=  \sum_{k=1}^{z} \frac{w_{k}}{\mathrm{Tr}\left[\varrho_{k}^{2}(t_{0})\right]} \mathrm{Tr}[O\varrho_{k}(t_{0})O^{\dag}\varrho_{k}(t_{\mathrm{f}})],
\label{eq.20}
\end{equation}
where $0\leqslant w_{k}\leqslant 1,\;\textstyle{\sum}_{k}w_{k}=1$. The actual state $\varrho_{k}(t_{\mathrm{f}})$ is a result of time evolution of the system with the initial state $\varrho_{k}(t_{0})$ at a given final time $t_{\mathrm{f}}$. Although one can assume $z=N^{2}$ \cite{Kallush_2006, Schulte-Herbruggen_2011}, in Ref. \cite{Goerz_2014} it has been argued that the following three initial states (i.e., $z=3$) suffice to determine suitable implementation of the desired unitary $O$:
\begin{align}
\varrho_{1}(t_{0}) =&\, (1/[N(N+1)])\textstyle{\sum_{i}} 2(N-i+1)  |i\rangle\langle i|,\label{eq.20-1}\\
\varrho_{2}(t_{0}) = &\, (1/N) \textstyle{\sum_{i,j}} |i\rangle\langle j|, \label{eq.20-2} \\
\varrho_{3}(t_{0}) = &\, (1/N) \textstyle{\sum_{i}} |i\rangle\langle i|,\label{eq.20-3}
\end{align}
where these initial states are defined in the optimization subspace such that other elements of these states are zero.

\subsection{Feature-based functionals}

Controlling (quantum) features is our principal object of interest. Specifically, we aim to control an open quantum system such that its dynamics at a given final time has a given feature. We focus on two particular features of a quantum process: ``purity'' and ``coherence.'' We want to study whether using associated measures of these features can yield a relative advantage compared to using process fidelity measures. This is an important question because two processes with high fidelity do not necessarily have similar quantum features \cite{PhysRevA.87.052136,bina2014drawbacks}.       

\subsubsection{Purity of a process}

Given a quantum process $\mathpzc{E}$, purity is defined as \cite{kong2016state}
\begin{equation}
 \mathpzc{P}(\mathpzc{E})=\langle\hskip-0.7mm\langle\varrho_{\mathpzc{E}}\vert \varrho_{\mathpzc{E}}\rangle\hskip-0.7mm\rangle = \dfrac{1}{N^{2}}\langle\hskip-0.7mm\langle\chi\vert \chi\rangle\hskip-0.7mm\rangle =: \mathpzc{P}(\chi). 
\label{eq.21}
\end{equation}
Equation (\ref{eq.5}) implies that this feature is basis independent and from Eq. \eqref{eq.6} it is bounded as $\mathpzc{P}(\chi)\leqslant 1$, while equality holds only when the dynamics is unitary. This feature provides a suitable measure to quantify the degree of unitarity of a quantum dynamics. As a result of time evolution of the system, purity of the initial unitary dynamics gradually diminishes and the dynamics traverses the hypersphere of the dynamical space $\mathfrak{D}_{S}$ due to the system-environment interaction (Fig. \ref{fig:ContSce}). 

\subsubsection{Coherence of a process}

The notion of coherence of states \cite{PhysRevLett.113.140401,PhysRevA.91.042120} has already been generalized to quantum dynamics/operations, and it has helped quantify the relevant properties for dynamics. Here closely follow Ref. \cite{xu2019coherence}. One first needs to consider a basis $\{|i\rangle\}$ and then define ``incoherent operations'' and ``incoherent superoperations'': a quantum operation or map $\mathpzc{E}_{\,\mathrm{incoh}}$ is incoherent if $\Upsilon[\mathpzc{E}_{\,\mathrm{incoh}}]=\mathpzc{E}_{\,\mathrm{incoh}}$, where $\Upsilon[\mathpzc{E}_{\,\mathrm{incoh}}]=\mathpzc{E}_{\,\mathrm{d}}\circ\mathpzc{E}_{\,\mathrm{incoh}}\circ\mathpzc{E}_{\,\mathrm{d}}$ and $\mathpzc{E}_{\,\mathrm{d}}[\varrho_{S}]=\textstyle{\sum}_{i}\langle i\vert\varrho_{S}\vert i\rangle\vert i\rangle\langle i\vert$ is a completely dephasing operation, with $\varrho_{S}$ being the density matrix of the system. We denote the set of all incoherent operations by $\mathscr{I}_{O}$. It can be shown that $\varrho_{\mathpzc{E}_{\,\mathrm{incoh}}}=(1/N)\textstyle{\sum}_{i,j=1}^{N}[\mathpzc{E}_{\,\mathrm{incoh}}]_{ii,jj}\vert ij\rangle\langle ij\vert$, where $[\mathpzc{E}_{\,\mathrm{incoh}}]_{ii,jj}=\langle j\vert\mathpzc{E}_{\,\mathrm{incoh}}[\vert i\rangle\langle i\vert]\vert j\rangle$. 

A superoperation $\Theta$ which maps an operation ($\mathpzc{E}$) to another operation ($\Theta[\mathpzc{E}]$) admits a Kraus representation as $\varrho_{\Theta[\mathpzc{E}]}=\textstyle{\sum}_{\alpha}\mathpzc{M}_{\alpha}\varrho_{\mathpzc{E}}\mathpzc{M}_{\alpha}^{\dagger}$, where $\varrho_{\mathpzc{E}}$ ($\varrho_{\Theta[\mathpzc{E}]}$) is the Choi matrix of $\mathpzc{E}$ ($\Theta[\mathpzc{E}]$) and $\textstyle{\sum}_{\alpha}\mathpzc{M}_{\alpha}^{\dagger}\mathpzc{M}_{\alpha}=\mathbbmss{I}$. A superoperation is called \textit{incoherent} if its Kraus operators have the forms as
$\mathpzc{M}_{\alpha}= \textstyle{\sum_{i,j}} \mathpzc{M}_{\alpha,(i,j)}\vert f(ij)\rangle\langle ij\vert, \quad f(ij)\in\big{\{}(i',j')\vert_{i',j'=1}^{N}\big{\}}$.
We denote the set of all incoherent superoperations by $\mathscr{I}_{S}$.

As argued in Ref. \cite{xu2019coherence}, the following necessary conditions can be conceived for any coherence measure $\mathpzc{C}(\mathpzc{E})$ for any operation $\mathpzc{E}$: (i) $\mathpzc{C}(\mathpzc{E})\geqslant 0$ for any quantum operation $\mathpzc{E}$ and $\mathpzc{C}(\mathpzc{E})=0$ for $\mathpzc{E}\in\mathscr{I}_{O}$; (ii)  monotonicity under incoherent superoperations on average. $\mathpzc{C}(\mathpzc{E})\geqslant \textstyle{\sum}_{\alpha}p_{\alpha}\mathpzc{C}(\mathpzc{E}_{\alpha})$ for any $\Theta\in\mathscr{I}_{S}$ with the Kraus operators $\{\mathpzc{M}_{\alpha}\}$, $p_{\alpha}=\mathrm{Tr}[\mathpzc{M}_{\alpha}\varrho_{\mathpzc{E}_{\alpha}}\mathpzc{M}_{\alpha}^{\dagger}]$ and $\varrho_{\mathpzc{E}_{\alpha}}=\mathpzc{M}_{\alpha}\varrho_{\mathpzc{E}}\mathpzc{M}_{\alpha}^{\dagger}/p_{\alpha}$; and (iii) Convexity. $\mathpzc{C}(\textstyle{\sum}_{\alpha}p_{\alpha}\mathpzc{E}_{\alpha})\leqslant \textstyle{\sum}_{\alpha}p_{\alpha}\mathpzc{C}(\mathpzc{E}_{\alpha})$ for any set of operations $\{\mathpzc{E}_{\alpha}\}$ and probability distribution $\{ p_{\alpha}\}$.

\begin{figure}[tp]
\includegraphics[width=0.6\linewidth]{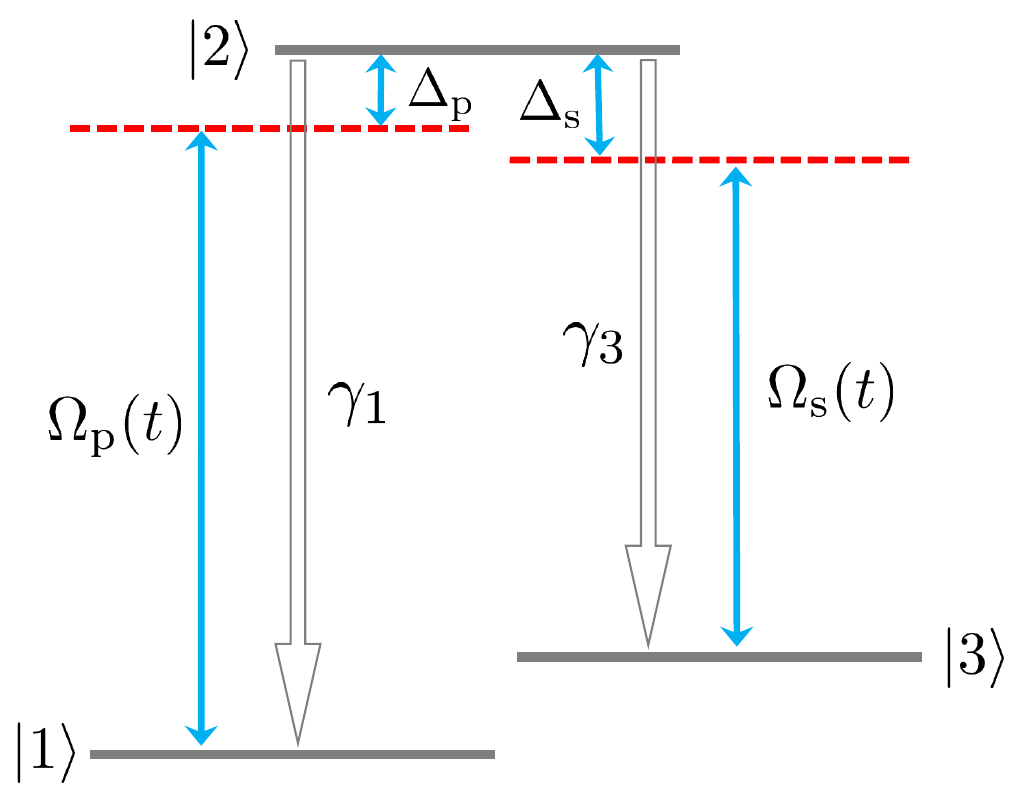}
\caption{Schematic of the $\Lambda$ configuration for a $3$-level quantum system. Two lower states $\ket{1}$ and $\ket{3}$ are coupled to an excited state $\ket{2}$, which decays with rates $\gamma_{\mathrm{1}}$ and $\gamma_{\mathrm{3}}$. The transition $\ket{1} \leftrightarrow \ket{2}$ is driven by a pump laser field with the Rabi frequency $\Omega_{\mathrm{p}}(t)$ and detuning $\Delta_{\mathrm{p}}$, and the transition $\ket{2} \leftrightarrow \ket{3}$ is driven by a Stokes laser field with the Rabi frequency $\Omega_{\mathrm{s}}(t)$ and detuning $\Delta_{\mathrm{s}}$.}
\label{fig:model}
\end{figure}

Reference \cite{xu2019coherence} has shown that if $\mathpzc{C}$ is a coherence measure of quantum states in the sense of Ref. \cite{PhysRevLett.113.140401}, then $\mathpzc{C}(\mathpzc{E}):=\mathpzc{C}(\varrho_{\mathpzc{E}})$ for quantum operation $\mathpzc{E}$ is a coherence measure for quantum operations. Accordingly, we consider the $\ell_{1}$ norm of the offdiagonal elements of the Choi-Jamiolkowski matrix $\varrho_{\mathpzc{E}}$ as a measure of quantum coherence of $\mathpzc{E}$ defined as
\begin{align}
\mathpzc{C}_{\ell_{1}}(\mathpzc{E}) &= \frac{1}{ N^{2} - 1}\textstyle{\sum_{\tilde{i} \neq \tilde{j}}} \vert \langle\hskip-0.7mm\langle\varrho_{\mathpzc{E}}\vert \tilde{C}_{(\tilde{i},\tilde{j})}\rangle \hskip-0.7mm \rangle \vert \label{eq.23}\\
&=\frac{1}{ N(N^{2} - 1)}\textstyle{\sum_{\tilde{i} \neq \tilde{j}}} \vert \langle\hskip-0.7mm\langle\mathpzc{U}^{\dag}{\chi}\mathpzc{U}\vert \tilde{C}_{(\tilde{i},\tilde{j})}\rangle\hskip-0.7mm\rangle\vert=:\mathpzc{C}_{\ell_{1}}(\chi).\nonumber
\end{align}
Note that the coherence measure is basis dependent and here we set 
$\{\tilde{C}_{(\tilde{i},\tilde{j})}=\vert \tilde{i}\rangle \langle \tilde{j} \vert\}_{\tilde{i}, \tilde{j}=1}^{N^{2}}$. In addition, $0 \leqslant \mathpzc{C}_{\ell_{1}} (\mathpzc{E}) \leqslant 1$ and the upper bound is saturated for an operation with a maximally coherent Choi-Jamiolkowski matrix (orange dots in Fig. \ref{fig:ContSce}). 

\section{Dynamical control of a Rydberg qutrit}
\label{sec.5}

To illustrate the feature-based control scheme, we focus on a $\Lambda$-type quantum system as in Fig. \ref{fig:model}, which can be realized, e.g., in the Rydberg atom $^{87}\mathrm{Rb}$ \cite{coleman2022exact}. This system with the energy levels $\{\ket{1}, \ket{2},\ket{3}\}$ constitutes a qutrit. It has been known that this system can be steered dynamically through two external laser fields which couple the states $\ket{1} \leftrightarrow \ket{2}$ (pump laser) as well as $\ket{2} \leftrightarrow \ket{3}$ (Stokes laser). By applying the rotating-wave approximation \cite{Shore_2011, goerz2015optimizing}, the time-dependent Hamiltonian of this system becomes
\begin{equation}
{H}_S(t) = \frac{\hbar}{2} \begin{pmatrix}
-2\Delta_{\mathrm{p}} & -\Omega_{\mathrm{p}}(t) & 0 \\
-\Omega_{\mathrm{p}}(t) & 0 & -\Omega_{\mathrm{s}}(t) \\
0 & -\Omega_{\mathrm{s}}(t) & -2\Delta_{\mathrm{s}}
\end{pmatrix},
\label{eq.24}
\end{equation}
where $\Delta_{\mathrm{p}}$ ($\Delta_{\mathrm{s}}$) is the detuning of the pump (Stokes) laser and $\Omega_{m}(t),\;m\in\{\mathrm{p},\mathrm{s}\}$ are the time-dependent Rabi frequencies of the driving external fields. Here we set $ \Delta_{\mathrm{p}} = \Delta_{\mathrm{s}} = 0.1\,\mathrm{MHz}$.

In this model the destructive effects of the system-environment interaction emerge as the two decaying quantum channels. The upper state $\ket{2}$ decays to two lower states with decay rates $\gamma_{1}$ and $\gamma_{3}$, respectively. Here we set these jump rates as $\gamma_{1} = \gamma_{3} = 0.1~\mathrm{MHz}$. Furthermore, the Lindblad jump operators in the $\Lambda$-system are defined as
\begin{align}
L_{2\to i} = |i\rangle \langle 2|,\quad i\in\{1,3\}.
\label{eq.25}
\end{align}

\begin{figure}[bp]
\includegraphics[width=0.45\textwidth]{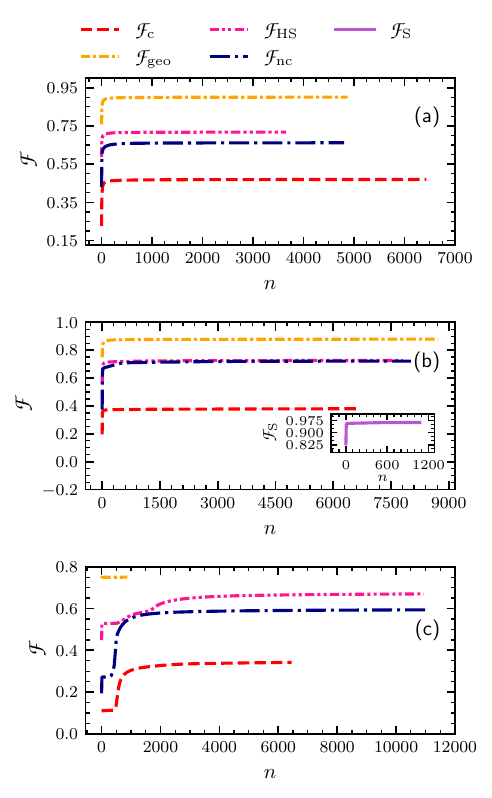}
\caption{Fidelity functional $\mathpzc{F}$ vs. the iteration number $n$ for the implementation of (a) $O_{\mathrm{phase},\varphi}$ at $t_{\mathrm{f}}=20\, \mu \mathrm{s}$, (b) $O_{\mathrm{QFT},q}$ at $t_{\mathrm{f}}=20\, \mu \mathrm{s}$, and (c) $O_{\mathrm{QFT},q}$ at $t_{\mathrm{f}}=200\, \mu \mathrm{s}$. Inset in (b) indicates the state-based overlap-based functional $\mathpzc{F}_{\mathrm{S}}$ (purple curve) for $O_{\mathrm{QFT},q}$ at $t_{\mathrm{f}}=20 ~\mu \mathrm{s}$.}
\label{fig:all-functionals}
\end{figure}

Having the time-dependent Hamiltonian \eqref{eq.24} and the Lindblad jump rates and operators \eqref{eq.25}, one can use Eq. \eqref{eq.8} to obtain a dynamical equation for the evolution of the driven $\Lambda$-system. Using this equation, the control fields are designed such that the final-time dynamics of the qutrit system resembles optimally a given target gate. We use the different fidelity functionals \eqref{eq.16} -- \eqref{eq.19} as measures of similarity between the target and simulated dynamics. The key point is that the generated dynamics using these functionals are similar in quantum features to the given target gate. The target process is
\begin{equation}
\left[ \Xi \right]_{\lambda \mu} \equiv \mathrm{Tr} [ O C_{\lambda}^{\dag} ] \,\mathrm{Tr} [ O C_{\mu}^{\dagger} ]^{*},\quad\lambda , \mu\in\{1,\ldots,N^{2}\},
\end{equation}
which, for example, we assume to correspond to the phase gate and the quantum Fourier transform (QFT) defined in the $\{|1\rangle,|2\rangle,|3\rangle\}$ basis as
\begin{gather}
O_{\mathrm{phase},\varphi}= \begin{pmatrix}1& 0 & 0\\ 0 & 1 & 0\\ 0 & 0& e^{i\varphi} \end{pmatrix},~~~~~~ O_{\mathrm{QFT},q} = \frac{1}{\sqrt{3}} 
\begin{pmatrix}
1 & 1 & 1 \\
1 & q & q^{2} \\
1 & q^{2} & q \\
\end{pmatrix},
\label{eq.26}
\end{gather}
respectively, where $\varphi = \pi$ and $q = e^{2\pi i / 3}$. 
Note that we set here the Gell-Mann basis matrices \cite{Bertlmann_2008} for a qutrit system as the orthonormal operator basis $\{ C_{\lambda}\}$. Through the feature-based optimal control, we also want to know how much one can increase purity or coherence of this process by applying appropriate external fields. For numerical implementations, we consider the following initial guess fields $f_{m}(t)$:
‎\begin{align}‎
&\Omega_{m}^{(0)}(t) = E^{0}_{m} f_{m}(t),\qquad \,\, m \in\{\mathrm{s}, \mathrm{p}\},\nonumber\\
&‎f_{m}(t)=[1-g-\cos(k_{m}\pi t/t_{\mathrm{f}})+g\cos(l_{m}\pi t/t_{\mathrm{f}})]/2‎,
‎\label{eq.27}‎
‎\end{align}
where $g=0.16$, $E^{0}_{\mathrm{p}} = 0.3 \,\mathrm{MHz}$, $E^{0}_{\mathrm{s}} = 1\, \mathrm{MHz}$,  and $k_{\mathrm{p}}=4,\; l_{\mathrm{p}}=8$ ($k_{\mathrm{s}}=2,\; l_{\mathrm{s}}=4$) for the pump (Stokes) laser. Since numerical algorithms often depend on initial guess points to reach a local optimal solution, we also test other shape functions such as Gaussian and sinusoidal,
‎\begin{align}‎
&f_{\mathrm{Gauss}}(t)=e^{-32[(t/t_{\mathrm{f}})-(1/2)]^{2}},‎\label{eq.28}\\
&f_{\mathrm{sin}}(t)=\sin^{2}(2\pi t/t_{\mathrm{f}}),
‎\label{eq.29}
‎\end{align}
respectively, to start the algorithm in some cases. In the following, we show that using educated or pre-optimized initial fields can improve performance of the optimization algorithm.

\subsection{Comparison of fidelity functionals}
\label{comparison-of-functionals}

Figures \ref{fig:all-functionals}a -- \ref{fig:all-functionals}c show the convex (red curve), nonconvex (blue curve), Hilbert-Schmidt (pink curve), and geometric (orange curve) fidelity functionals vs. the iteration number $n$ to simulate the QFT and phase gates at final times $t_{\mathrm{f}}=20\,\mu \mathrm{s}$ and $t_{\mathrm{f}}=200\,\mu \mathrm{s}$. The Krotov algorithm guarantees to reach an optimal solution for total objective functionals $\mathpzc{J}$'s monotonically. It is also evident from these figures that the values of the final-time fidelity functionals $\mathpzc{F}$'s increase monotonically with respect to the iteration number. 

At first glance it seems that $\mathpzc{F}_{\mathrm{geo}}$ outperforms the other measures. However, for a fair quantitative comparison, we first consider the dynamics resulting from the optimization of a fidelity functional, shown by ${\mathpzc{F}}_{k}$, and then by incorporating the optimal fields we calculate the other fidelities $\mathpzc{F}_{l}$ ($l\in\{\mathrm{c},\mathrm{nc},\mathrm{HS},\mathrm{geo}\}$) between this dynamics and the desired unitary gate. We use the notation $\mathpzc{F}_{l}\big|_{{\mathpzc{F}}_{k}}$ for this quantity and $\mathpzc{F}_{l}\big|_{{\mathpzc{F}}_{l}}={\mathpzc{F}}_{l}$. If $\mathpzc{F}_{l}\big|_{{\mathpzc{F}}_{k}}\leqslant{\mathpzc{F}}_{l}$ and $\mathpzc{F}_{k}\big|_{{\mathpzc{F}}_{l}}\geqslant{\mathpzc{F}}_{k}$ for $l\neq k$, then the final-time functional $\mathpzc{F}_{l}$ has outperformed $\mathpzc{F}_{k}$ in optimization. The same expression holds for the fidelity functional $\mathpzc{F}_{k}$. 

\begingroup
\squeezetable
\begin{table}[tp]
\caption{$\mathpzc{F}_{l}\big|_{{\mathpzc{F}}_{k}}$ with $k,l\in\{\mathrm{n},\mathrm{nc},\mathrm{HS},\mathrm{geo}\}$, where $\mathpzc{F}_{l}$ (column) is the fidelity between the dynamics obtained from the optimization of $\mathpzc{F}_{k}$ (row) and $O_{\mathrm{phase},\varphi}$ at $t_{\mathrm{f}} = 20 ~\mu \mathrm{s}$.}
\begin{ruledtabular}
\begin{tabular}{lcccc}
optimization functional & \multicolumn{4}{c}{functional} \\
  \cmidrule(lr){2-5}
 & $\mathpzc{F}_{\mathrm{c}}$ & $\mathpzc{F}_{\mathrm{nc}}$ & $\mathpzc{F}_{\mathrm{HS}}$ & $\mathpzc{F}_{\mathrm{geo}}$ \\
\midrule
$\mathpzc{F}_{\mathrm{c}}$  & $0.470$  & $0.662$ & $0.718$ & $0.9$ \\
$\mathpzc{F}_{\mathrm{nc}}$ & $0.469$ & $0.662$  & $0.718$ & $0.9$ \\
$\mathpzc{F}_{\mathrm{HS}}$ & $0.467$  & $0.661$  & $0.718$ & $0.9$ \\
$\mathpzc{F}_{\mathrm{geo}}$ & $0.469$  & $0.662$ & $0.718$  & $0.9$  \\
\end{tabular}
\end{ruledtabular}
\label{table.1}
\end{table}
\endgroup

\begin{figure}[tp]
\includegraphics[width=0.45\textwidth]{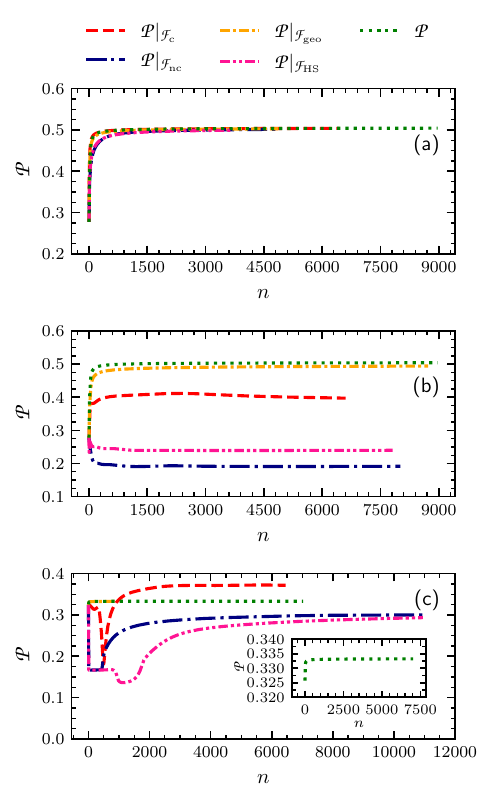}
\caption{Purity of the process matrix obtained from the optimization of the nonconvex overlap-based $\mathpzc{P}|_{\mathpzc{F}_{\mathrm{nc}}}$ (blue dashed dot curve), convex overlap-based $\mathpzc{P}|_{\mathpzc{F}_{\mathrm{c}}}$ (red dashed curve), geometric $\mathpzc{P}|_{\mathpzc{F}_{\mathrm{geo}}}$ (orange dashed dot curve), and Hilbert-Schmidt $\mathpzc{P}|_{\mathpzc{F}_{\mathrm{HS}}}$ (pink dashed double dot curve) vs. iteration number $n$ for the implementation of (a) $O_{\mathrm{phase},\varphi}$ at $t_{\mathrm{f}}=20 ~\mu \mathrm{s}$, (b) $O_{\mathrm{QFT},q}$ at $t_{\mathrm{f}}=20 ~\mu \mathrm{s}$, and (c) $O_{\mathrm{QFT},q}$ at $t_{\mathrm{f}}=200 ~\mu \mathrm{s}$. The purity functionals $\mathpzc{P}$ vs. $n$ have been shown by green dot curves resulting from optimization at times associated with these figures. Inset (c) represents the purity optimization at $t_{\mathrm{f}}=200 ~\mu \mathrm{s}$.}
\label{fig:all-purities}
\end{figure}

\begin{figure}[tp]
\includegraphics[width=0.45\textwidth]{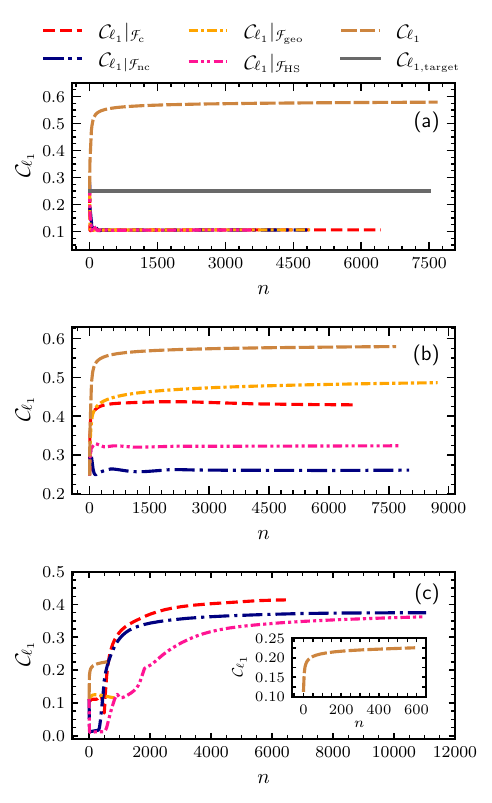}
\caption{Coherence obtained from the optimization of the nonconvex overlap-based $\mathpzc{C}_{\ell_{1}}\big|_{\mathpzc{F}_{\mathrm{nc}}}$ (blue dashed dot curve), convex overlap-based $\mathpzc{C}_{\ell_{1}}\big|_{\mathpzc{F}_{\mathrm{c}}}$ (red dashed curve), geometric $\mathpzc{C}_{\ell_{1}}\big|_{\mathpzc{F}_{\mathrm{geo}}}$ (orange dashed dot curve), Hilbert-Schmidt $\mathpzc{C}_{\ell_{1}}\big|_{\mathpzc{F}_{\mathrm{HS}}}$ (pink dashed double dot curve), and the coherence of the target process matrix $\mathpzc{C}_{\ell_{1},{\mathrm{target}}}$ (black curve) vs. iteration number $n$ for the implementation of (a) $O_{\mathrm{phase},\varphi}$ at $t_{\mathrm{f}}=20 ~\mu \mathrm{s}$, (b) $O_{\mathrm{QFT},q}$ at $t_{\mathrm{f}}=20 ~\mu \mathrm{s}$, and (c) $O_{\mathrm{QFT},q}$ at $t_{\mathrm{f}}=200 ~\mu \mathrm{s}$. The coherence functionals $\mathpzc{C}_{\ell_{1}}$ vs. $n$ have been shown by brown dashed curves resulting from optimization at times associated with these figures. Inset of (c) indicates coherence optimization at $t_{\mathrm{f}}=200 ~\mu \mathrm{s}$.}
\label{fig:all-coherence}
\end{figure}

By inspecting Table \ref{table.1}, we conclude that there is no difference between the final-time functionals $\mathpzc{F}_{l},\; l\in\{\mathrm{c},\mathrm{nc},\mathrm{HS},\mathrm{geo}\}$ in simulating $O_{\mathrm{phase},\varphi}$ at $t_{\mathrm{f}}=20\;\mu \mathrm{s}$. However, we need to compare the quality of the dynamics produced by these functionals for a better conclusion, referred to as a \textit{qualitative} comparison. Figures \ref{fig:all-purities}a and \ref{fig:all-coherence}a indicate the purity ($\mathpzc{P}\big|_{\mathpzc{F}_{k}}$) and the coherence ($\mathpzc{C}_{\ell_{1}}\big|_{\mathpzc{F}_{k}}$) of the dynamics produced through optimization of the fidelity functional $\mathpzc{F}_{k},\;k\in\{\mathrm{c},\mathrm{nc},\mathrm{HS},\mathrm{geo}\}$ with the target gate $O_{\mathrm{phase},\varphi}$ and $t_{\mathrm{f}}=20\,\mu \mathrm{s}$ vs. iteration number. All purity and coherence functionals converge to the same value so that $\mathpzc{P}\big|_{\mathpzc{F}_{k}} \approx 0.5$ and $\mathpzc{C}_{\ell_{1}}\big|_{\mathpzc{F}_{k}} \approx 0.105$ for all $k$. Noting $\mathpzc{P}_{\mathrm{target}}=1$, we observe that using the functionals has not affected considerably the amount of optimal purity. 

Table \ref{table.2} shows $\mathpzc{F}_{l}\big|_{\mathpzc{F}_{k}}$ for the desired $O_{\mathrm{QFT},q}$ gate and $t_{\mathrm{f}}=20\;\mu \mathrm{s}$. By comparing $\mathpzc{F}_{l}\big|_{\mathpzc{F}_{k}}$s, one concludes that the convex fidelity $\mathpzc{F}_{\mathrm{c}}$ has led to a relatively better result than the others; e.g., $\mathpzc{F}_{\mathrm{c}}-\mathpzc{F}_{\mathrm{c}}\big|_{\mathpzc{F}_{\mathrm{geo}}}\approx 0.008$ and $\mathpzc{F}_{\mathrm{geo}} \approx \mathpzc{F}_{\mathrm{geo}}\big|_{\mathpzc{F}_{\mathrm{c}}}$. However, the geometric fidelity has led to a better result compared to the others measures; $\mathpzc{P}\big|_{\mathpzc{F}_{\mathrm{geo}}}-\mathpzc{P}\big|_{\mathpzc{F}_{\mathrm{c}}}\approx 0.097 $ and $\mathpzc{C}_{\ell_{1}}\big|_{\mathpzc{F}_{\mathrm{geo}}}-\mathpzc{C}_{\ell_{1}}\big|_{\mathpzc{F}_{\mathrm{c}}}\approx 0.057$ (Figs. \ref{fig:all-purities}b and \ref{fig:all-coherence}b). Thus, we observe that the geometric fidelity (and next the convex overlap-based) functional has a comparative advantage over the other fidelities studied here.

We make a similar analysis for simulating $O_{\mathrm{QFT},q}$ at $t_{\mathrm{f}}=200\;\mu \mathrm{s}$. In particular, we want to obtain more information on the role of the predetermined final time in improving performance of the optimization algorithm via different fidelity functionals. The values of $\mathpzc{F}_{l}\big|_{\mathpzc{F}_{k}}$ with $l,k\in\{\mathrm{c},\mathrm{nc},\mathrm{HS},\mathrm{geo}\}$ have been listed in Table \ref{table.3}. 

\begingroup
\squeezetable
\begin{table}[bp]
\caption{$\mathpzc{F}_{l}\big|_{{\mathpzc{F}}_{k}}$ with $k,l\in\{\mathrm{n},\mathrm{nc},\mathrm{HS},\mathrm{geo}\}$, where $\mathpzc{F}_{l}$ (column) is the fidelity between the dynamics obtained from the optimization of $\mathpzc{F}_{k}$ (row) and $O_{\mathrm{QFT},q}$ at $t_{\mathrm{f}} = 20 ~\mu \mathrm{s}$.}
\begin{ruledtabular}
\begin{tabular}{lcccc}
optimization functional & \multicolumn{4}{c}{functional} \\
\cmidrule(lr){2-5}
 & $\mathpzc{F}_{\mathrm{c}}$ & $\mathpzc{F}_{\mathrm{nc}}$ & $\mathpzc{F}_{\mathrm{HS}}$ & $\mathpzc{F}_{\mathrm{geo}}$ \\
\midrule
${\mathpzc{F}}_{\mathrm{c}}$  & $0.380$  & $0.602$  & $0.68$  & $0.855$ \\
${\mathpzc{F}}_{\mathrm{nc}}$  & $0.316$ & $0.722$  & $0.72$  & $0.748$ \\
${\mathpzc{F}}_{\mathrm{HS}}$  & $0.345$  & $0.706$  & $0.725$ & $0.775$ \\
${\mathpzc{F}}_{\mathrm{geo}}$ & $0.372$  & $0.528$  & $0.62$  & $0.88$  \\
\end{tabular}
\end{ruledtabular}
\label{table.2}
\end{table}
\endgroup

For this optimization we observe that $\mathpzc{F}_{\mathrm{geo}}$ underperforms in simulating this unitary dynamics, while $\mathpzc{F}_{\mathrm{c}}$, $\mathpzc{F}_{\mathrm{nc}}$, and $\mathpzc{F}_{\mathrm{HS}}$ perform almost similarly. However, $\mathpzc{F}_{\mathrm{HS}}$ underperforms $\mathpzc{F}_{\mathrm{nc}}$ and has no advantage over the convex fidelity. It seems that the quantitative differences between these final-time functionals become more apparent as the final time $t_{\mathrm{f}}$ increases. Thus, based on their quantitative values, the nonconvex overlap-based and the convex functional are more suitable, noting also that their difference is negligible, $\mathpzc{F}_{\mathrm{c}}-\mathpzc{F}_{\mathrm{c}}\big|_{\mathpzc{F}_{\mathrm{nc}}}\approx 0.017$ and $\mathpzc{F}_{\mathrm{nc}}-\mathpzc{F}_{\mathrm{nc}}\big|_{\mathpzc{F}_{\mathrm{c}}}\approx 0.032$. For a definitive conclusion, we need to evaluate performance of each functional for high-quality gate implementation. Purity and coherence of the dynamics $\chi^{(n)}(t_{\mathrm{f}})$ resulting from the optimization of these functionals vs. iteration number has been represented in Figs. \ref{fig:all-purities}c and \ref{fig:all-coherence}c, respectively. Purity of the dynamics produced through optimization of the convex overlap-based functional is higher than the other functionals. For example, the difference between these quantities and their subsequent larger values belonging to the geometric and nonconvex functionals are $\mathpzc{P}\big|_{\mathpzc{F}_{\mathrm{c}}}-\mathpzc{P}\big|_{\mathpzc{F}_{\mathrm{geo}}}\approx 0.039$ and $\mathpzc{C}_{\ell_{1}}\big|_{\mathpzc{F}_{\mathrm{c}}}-\mathpzc{C}_{\ell_{1}}\big|_{\mathpzc{F}_{\mathrm{nc}}}\approx 0.04 $. The convex functional hence outperforms the others in generating a unitary dynamics with a relatively high quality. 

\begingroup
\squeezetable
\begin{table}[bp]
\caption{$\mathpzc{F}_{l}\big|_{{\mathpzc{F}}_{k}}$ with $k,l\in\{\mathrm{c},\mathrm{nc},\mathrm{HS},\mathrm{geo}\}$, where $\mathpzc{F}_{l}$ (column) is the fidelity between the dynamics obtained from the optimization of $\mathpzc{F}_{k}$ (row) and $O_{\mathrm{QFT},q}$ at $t_{\mathrm{f}} = 200 ~\mu \mathrm{s}$.}
\begin{ruledtabular}
\begin{tabular}{lcccc}
optimization functional & \multicolumn{4}{c}{functional} \\
  \cmidrule(lr){2-5}
 & $\mathpzc{F}_{\mathrm{c}}$ & $\mathpzc{F}_{\mathrm{nc}}$ & $\mathpzc{F}_{\mathrm{HS}}$ & $\mathpzc{F}_{\mathrm{geo}}$ \\
\midrule
$\mathpzc{F}_{\mathrm{c}}$  & $0.342$  & $0.562$  & $0.657$  & $0.837$ \\
$\mathpzc{F}_{\mathrm{nc}}$ & $0.325$ & $0.594$  & $0.675$  & $0.802$ \\
$\mathpzc{F}_{\mathrm{HS}}$  & $0.316$ & $0.584$  & $0.670$ & $0.796$ \\
$\mathpzc{F}_{\mathrm{geo}}$ & $0.111$  & $0.193$  & $0.445$ & $0.750$ \\
\end{tabular}
\end{ruledtabular}
\label{table.3}
\end{table}
\endgroup

To sum up so far, through a fair quantitative and qualitative comparison of various fidelity functionals, it seems that the convex overlap-based functional ($\mathpzc{F}_{\mathrm{c}}$) has a relative advantage over the others to simulate a target quantum gate with high fidelity at a given final time. However, in most applications investigated in the literature so far, the state-based fidelity has often been used as the final-time functional \cite{Schulte-Herbruggen_2011, Floether_2012, Goerz_2014, PhysRevLett.111.200401}. 
Thus, the comparison of such fidelities with the preferred functional, i.e., convex overlap-based functional, is an imperative task. Here we consider the reduced basis-dependent functional $\mathpzc{F}_{\mathrm{S}}$ as a state-dependent final time functional [Eq. \eqref{eq.20}] with only three input states [Eqs. \eqref{eq.20-1} -- \eqref{eq.20-3}]. Such reduction in inputs is independent of the Hilbert space dimension of the system and then reduces considerably the computational cost of the numerics \cite{Goerz_2014}. But it has remained to be seen whether this computational advantage is also accompanied by a relative advantage in constructing a target quantum gate. 

Inset of Fig. \ref{fig:all-functionals}b shows $\mathpzc{F}_{\mathrm{S}}$ vs. the iteration number for the target gate $O_{\mathrm{QFT},q}$ at $t_{\mathrm{f}}=20\;\mu \mathrm{s}$. The state-based fidelity eventually converges to the value $0.963$. The results of the optimizations with $\mathpzc{F}_{\mathrm{c}}$ and $\mathpzc{F}_{\mathrm{S}}$ have been indicated in Table \ref{table.4}. Since $\mathpzc{F}_{\mathrm{c}}>{\mathpzc{F}}_{\mathrm{c}}\big|_{{\mathpzc{F}}_{\mathrm{S}}}$ and $\mathpzc{F}_{\mathrm{S}}>{\mathpzc{F}}_{\mathrm{S}}\big|_{{\mathpzc{F}}_{\mathrm{c}}}$, then there is no preference between them quantitatively. However, by comparing the quantum features of the final-time dynamics, we observe that purity increases by $\approx 43\%$ from its initial value by using the convex functional, but this quantity does not change for the state-dependent functional $\mathpzc{F}_{\mathrm{S}}$. Coherence of the final-time dynamics also increases by $\approx 73\%$ and $\approx 10\%$, respectively from its initial value through the optimization of $\mathpzc{F}_{\mathrm{c}}$ and $\mathpzc{F}_{\mathrm{S}}$. The convex functional $\mathpzc{F}_{\mathrm{c}}$ hence outperforms the functional $\mathpzc{F}_{\mathrm{S}}$ in reproducing $O_{\mathrm{QFT},q}$ with a relatively higher quality.   

\begingroup
\squeezetable
\begin{table}[bp]
\caption{Optimization of the convex overlap-based $\mathpzc{F}_{\mathrm{c}}$ and state-based $\mathpzc{F}_{\mathrm{S}}$ at $ t_{\mathrm{f}} = 20 ~\mu \mathrm{s}$ with $O_{\mathrm{QFT},q}$ as the target gate.}
\label{process-table}
\begin{ruledtabular}
\begin{tabular}{lccc}
functional & initial value & optimal value (by $\mathpzc{F}_{\mathrm{c}}$) & optimal value (by $\mathpzc{F}_{\mathrm{S}}$) \\
\midrule
$\mathpzc{F}_{\mathrm{c}}$ & $0.196$ & $0.380$ & $0.290$ \\
$\mathpzc{P}$ & $0.278$ & $0.397$ &  $0.278$\\
$\mathpzc{C}_{\ell_1}$ & $0.248$ & $0.43$ & $0.273$ \\
$\mathpzc{F}_{\mathrm{S}}$  & $0.831$ & $0.881$ & $0.963$ \\
\end{tabular}
\end{ruledtabular}
\label{table.4}
\end{table}
\endgroup

\subsection{Feature-based optimal control}
\label{sec.5-subsec.b}

Having purity and coherence of a quantum dynamics enables us to design a control scheme for increasing the quantum features as much as possible despite the destructive effects of the environment. Indeed, the main objective of this feature- and process-based control is not to approach the final dynamics to a target gate. Rather, we want to design the appropriate external fields to guide the dynamics at $t_{\mathrm{f}}$ to a particular subset of $\mathfrak{D}_{S}$ labeled by some quantum features, e.g., dynamics with the maximum purity (unitary dynamics) or maximum coherence (green and orange dashed curves of Fig. \ref{fig:ContSce}). 

\begin{figure}[tp]
\includegraphics[width=0.43\textwidth]{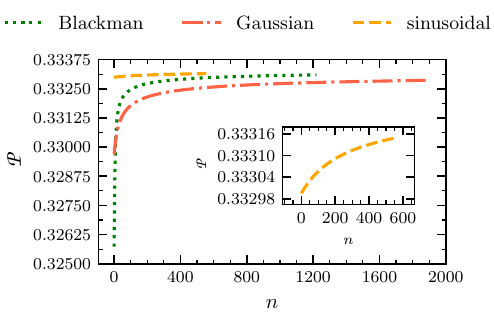}
\caption{Purity $\mathpzc{P}$ vs. iteration number $n$ for $t_{\mathrm{f}}=200 ~\mu \mathrm{s}$, with the Blackman \eqref{eq.27} (green dot curve), sinusoidal  \eqref{eq.28} (orange dashed curve), and Gaussian \eqref{eq.29} (red dashed curve) shape functions as the initial guess fields. Inset represents optimization at $t_{\mathrm{f}}=200 ~\mu \mathrm{s}$ with the sinusoidal shape functions as the guess fields. The convergence condition in all optimizations is $\Delta \mathpzc{J} < 10^{-7}$.}
\label{fig:purity-shapes}
\end{figure}

Since the $\ell_{1}$ norm coherence [Eq. \eqref{eq.23}] is a nonconvex function of the process matrix, the Krotov algorithm is a suitable option to solve the above optimization problem. The green dotted and brown dashed curves in Figs. \ref{fig:all-purities} and \ref{fig:all-coherence} show the purity and $\ell_{1}$ norm of coherence vs. iteration number. The purity eventually approaches $\approx$ 0.504 at the final time $t_{\mathrm{f}}=20\;\mu \mathrm{s}$ which coincides with $\mathpzc{P}\big|_{\mathpzc{F}_{k}},\;k\in\{\mathrm{c},\mathrm{nc},\mathrm{HS},\mathrm{geo}\}$ for $O_{\mathrm{phase},\varphi}$ as a desired gate (Fig. \ref{fig:all-purities}a). However, the optimized purity is higher than the purity of dynamics obtained from optimization of all fidelity functionals with quantum Fourier transform as a target gate (Fig. \ref{fig:all-purities}b). This difference becomes more considerable by considering $\mathpzc{C}_{\ell_{1}}$ as the final time functional for the predetermined time $t_{\mathrm{f}}=20\;\mu \mathrm{s}$ (Figs. \ref{fig:all-coherence}a and \ref{fig:all-coherence}b). The optimized dynamics has a coherence $\mathpzc{C}_{\ell_{1}}\approx 0.57$ which is higher than coherence of dynamics obtained from optimization of all fidelity functionals, i.e., $\mathpzc{C}_{\ell_{1}}\big|_{\mathpzc{F}_{k}},\; k\in\{\mathrm{c},\mathrm{nc},\mathrm{HS},\mathrm{geo}\}$, with $O_{\mathrm{phase},\varphi}$ and $O_{\mathrm{QFT},q}$ as the target unitary dynamics. 

\begin{figure}[tp]
\includegraphics[width=0.45\textwidth]{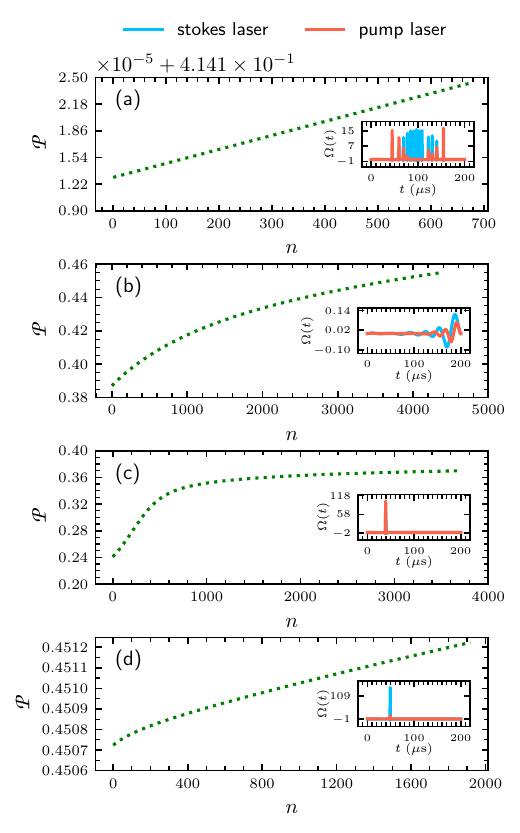}
\caption{Purity $\mathpzc{P}$ vs. iteration number $n$ at $t_{\mathrm{f}}=200 ~\mu \mathrm{s}$ with (a) the pre-optimized fields obtained from optimization of the nonconvex overlap-based functional at the same time for $O_{\mathrm{phase},\varphi}$ as the target gate,  the pre-optimized and rescaled fields obtained from optimization of purity functional at earlier times (b) $t_{\mathrm{f}} = 5 ~\mu \mathrm{s}$, (c) $t_{\mathrm{f}} = 15 ~\mu \mathrm{s}$, and (d) $t_{\mathrm{f}} = 20 ~\mu \mathrm{s}$, as the initial guess fields. Insets indicate the educated pump (red curve) and Stokes (blue curve) fields.}
\label{fig:preoptimized-purities}
\end{figure}

Purity of the dynamics varies only negligibly as a result of optimization of the purity functional at $t_{\mathrm{f}}=200\;\mu \mathrm{s}$ (inset of Fig. \ref{fig:all-purities}c). The optimal value is lower than $\mathpzc{P}\big|_{\mathpzc{F}_{c}}$ due to the optimization of $\mathpzc{F}_{c}$ with the target gate $O_{\mathrm{QFT},q}$ and matches $\mathpzc{P}\big|_{\mathpzc{F}_{\mathrm{geo}}}$. This situation gets worse when we optimize the functional $\mathpzc{C}_{\ell_{1}}$ at the given final time $t_{\mathrm{f}}=200\;\mu \mathrm{s}$ (Fig. \ref{fig:all-coherence}c and its inset). The final value of the $\ell_{1}$ norm of coherence, i.e., $\approx$ 0.226, is lower than the coherence obtained from the optimization of the three fidelity functionals $\mathpzc{P}\big|_{\mathpzc{F}_{k}},\; k\in\{\mathrm{c},\mathrm{nc},\mathrm{HS}\}$, with $O_{\mathrm{QFT},q}$ as the target gate. 

At first sight, it might be perceived that by increasing the final time the optimization of quantum features does not lead to better dynamics compared to using the fidelity functionals. However, it should be noted that most optimization algorithms depend on initial guess points \cite{Werschnik_2007, Floether_2012, PhysRevA.91.043401, Goerz20155}. Hence a suitable choice to start the numerical algorithm may give rise to a higher (local) optimal point. For this reason and as an example, we consider other initial guess fields with different shape functions including Gaussian and sinusoidal functions [Eqs. \eqref{eq.28} and \eqref{eq.29}], instead of the Blackman shape to optimize purity at $t_{\mathrm{f}}=200\;\mu \mathrm{s}$. The optimization results with these starting points are shown in Fig. \ref{fig:purity-shapes}. It is observed that all three optimization processes converge almost to the same value $\approx 0.333$. 

The optimal value of purity can be improved by using the pre-optimized initial fields (Figs. \ref{fig:preoptimized-purities}a -- \ref{fig:preoptimized-purities}d). For example, we use the optimal fields resulting from the optimization of $\mathpzc{F}_{\mathrm{nc}}$, for the target gate $O_{\mathrm{QFT},q}$ and $t_{\mathrm{f}}=200\;\mu \mathrm{s}$, as the guess fields to optimize purity at the same final time (Fig. \ref{fig:preoptimized-purities}a). Although the optimal value of purity increases to $\approx 0.414$, this value does not change considerably during the optimization. Alternatively, we can employ the pre-optimized and rescaled initial fields to start the main optimization algorithm. To make such \textit{educated} fields, we first obtain optimal fields resulting from the optimization of purity at a different final time and then rescale them to the original final time. Figures \ref{fig:preoptimized-purities}b, \ref{fig:preoptimized-purities}c, and \ref{fig:preoptimized-purities}d show the purity functional vs. iteration number for intermediate times $t_{\mathrm{f}}=5\;\mu \mathrm{s}$, $t_{\mathrm{f}}=15\;\mu \mathrm{s}$, and $t_{\mathrm{f}}=20\;\mu \mathrm{s}$, respectively, and the original final time $t_{\mathrm{f}}=200\;\mu \mathrm{s}$. As seen from these figures, optimization improves quantitatively by setting the intermediate final time as $t_{\mathrm{f}}=5\;\mu \mathrm{s}$ such that the optimal value of purity increases by $\approx 38\%$ compared to the original optimization with the Blackman shape function at $t_{\mathrm{f}}=200\;\mu \mathrm{s}$. In fact, presence of different optimal local points indicates a nonsmooth topology for purity, which has been previously observed for overlap- and state-based fidelity measures \cite{Floether_2012, PhysRevLett.106.120402, PhysRevA.91.043401, HSIEH200877, Moore_2012}.

\section{Conclusions and outlook}
\label{sec.6}

We have investigated an optimal control framework for generating quantum processes which have some target quantum features. This is distinct from ordinary problems in quantum control in which the goal is often to generate a given target process, whereas here one may have several processes. We have considered the setting where an open quantum system undergoes a Markovian evolution in its environment, and the system has been coherently controlled by applying optimal control fields. We have investigated how choosing objective functional for the control problem affects the performance of the optimal control scheme. In particular, we have compared various process-based (and also a state-based) fidelity functionals vs. the purity and coherence functionals---as the features of interest. We have illustrated our feature-based optimal process control through a Rydberg qutrit system interacting with the environment. In this model, we have demonstrated the relative advantage of the feature-based functionals compared to fidelity-based functionals in some regimes. In particular, we have shown that by pre-optimization with respect to the fidelity functionals one can improve performance the feature-based optimization through providing educated initial guess fields for the optimization algorithm.

The control scheme presented here is based on manipulating the dynamics coherently to achieve the desired target process. One can extend this scheme to full dissipative control of an open-system dynamics. In addition, in this approach, the focus is on increasing the quantum features in a predetermined final time. One can also generalize this scheme to control these features during the whole time evolution of the system, as in the quantum spline problem for closed systems \cite{PhysRevLett.109.100501,doi:10.1142/S0219887818501475}.\\

\textit{Acknowledgments.}---V.R. gratefully acknowledges the comments by C. P. Koch on the Krotov algorithm. This work was supported by the Iran National Science Foundation (INSF) under Grant No. $4030102$ and Sharif University of Technology's Vice President for Research and Technology under Grant No. $\mathrm{QST}4040202$.

\bibliography{refs}
\end{document}